\documentstyle[prl,aps,epsf,psfig]{revtex}
\begin{document}
\twocolumn[
\hsize\textwidth\columnwidth\hsize\csname@twocolumnfalse\endcsname
\draft
\bibliographystyle{prsty}

\title{Many-body spin related phenomena in ultra-low-disorder quantum
wires}
\author{D. J. Reilly, G. R. Facer,\cite{geoff} A. S. Dzurak, B. E. Kane,\cite{bruce}
R. G. Clark, P. J. Stiles,\cite{phil} J. L. O'Brien and N. E. Lumpkin}

\address{ Semiconductor Nanofabrication Facility, University of New South
Wales, Sydney \space 2052, Australia}

\author{L. N. Pfeiffer and K. W. West}
\address{Bell Laboratories, Lucent Technologies, Murray Hill, New Jersey
\space 07974, USA\\}
\maketitle
\begin{abstract}
\noindent
Zero length quantum wires (or point contacts) exhibit unexplained conductance structure
close to  $0.7 \times 2e^2/h$ in the absence of an applied  
magnetic field. We have studied the density- and temperature-dependent conductance of 
ultra-low-disorder GaAs/AlGaAs
quantum
wires with nominal lengths $l$=0 and $2 \mu m$, fabricated from structures free
of the disorder associated with modulation doping. 
In a direct comparison we observe structure near $0.7 \times 2e^2/h$ for $l$=0 whereas the $l=2\mu m$ 
wires show structure evolving with
increasing electron density to $0.5 \times 2e^2/h$ in {\it zero} magnetic field, the value expected for an ideal
spin-split sub-band. Our results suggest the dominant mechanism 
through which electrons interact can be strongly affected by the length of the 1D region.
\end{abstract}     

\pacs{ 73.61.-r, 73.23.Ad, 73.61.Ey III-V}

]

Quantum wires have been used extensively to study ballistic  transport in one dimension (1D) where the
conductance is quantised in units of $2e^2/h$ \cite{Wharam_QPC1st,vanWees_QPC}.
 This
result is 
well explained by considering the
allowed 
energies of a non-interacting
electron gas confined to 1D,  where the factor 
of 2 is due to spin degeneracy. Electron interaction effects in 1D have been considered for some time,
involving
models \cite{Luttinger} which go beyond the conventional Fermi liquid picture. Such 
correlated electron models have been applied to quantum wire systems \cite{KandF_LLPRL}
and recent experimental studies \cite{Tarucha_quant,Yacoby_QWR} have investigated their predictions. Although
recent theories have
considered the effect of weak disorder on correlation effects \cite{Maslov_endNov}, it is generally accepted
that low-disorder nanostructures are necessary for such investigations. 

Low-disorder quantum point contacts (which are quantum wires of length $l$ = 0) 
formed in GaAs/AlGaAs
heterostructures exhibit unexplained conductance structure close 
to $0.7 \times 2e^2/h$ in the absence of a  
magnetic field \cite{Thomas_spin1st,Thomas_int,BK_QWAPL,Kristensen,Tsch1}.
Studies by
Thomas {\em et al.} \cite{Thomas_spin1st}  suggest that the structure is a
manifestation of
electron--electron interactions involving spin. The continuous 
evolution of the $0.7 \times 2e^2/h$ structure into a Zeeman spin--split 
conductance plateau  with the application of an in-plane
magnetic field, together with enhancement of the $g$-factor for lower 1D channels, 
is consistent with this interpretation \cite{Thomas_spin1st}. 

In this article we present transport data for 1D systems
free from the disorder 
associated with modulation doped heterostructures, including strong evidence for  spin related many-body 
effects in long 1D regions. We 
find conductance structure comparable to Thomas' in
our zero length wires,
while our  2 $\mu m $ quantum wire
exhibits plateau--like structure near
$0.5 \times 2e^2/h$ in zero magnetic field, the value expected for an ideal spin-split level.  

Our results are suggestive of an interpretation in which spin splitting in zero magnetic 
field is only fully resolved in long 1D regions, perhaps above a critical length scale. 
This does not explain why structure in short constrictions consistently occurs near $0.7 \times 2e^2/h$. A
clue may be found in recent theories \cite{flambaum,coloumb} which consider the possibility of a 
feature at $0.75 \times 2e^2/h$
due to a splitting between the one singlet and three triplet states where electron pairs (attractive interaction)
dominate transport. As the length of the 1D region is increased 
it is suggested \cite{flambaum} that the dominant many-body interaction can alter and if spontaneous spin 
polarisation occurs, 
a principal feature at $0.5 \times 2e^2/h$ would be observed, perhaps with some remanent weak structure close
to $0.75 \times 2e^2/h$. 

The study of correlated electron states requires devices with
ultra-low-disorder
since such states are expected to be easily destroyed by disorder and may be
masked by other effects associated with localisation.  We have developed
a novel GaAs/AlGaAs layer structure which avoids the major random potential
present in conventional HEMT
devices by using epitaxially grown gates to produce an enhancement mode FET
\cite{BK_FETnano}. These devices are
advantageous for the study of 1D interacting systems because they eliminate
the need for a dopant layer in the AlGaAs adjacent to the 2DEG, thus greatly
reducing disorder while allowing  the electron density in the
2DEG to be varied over a large range. The electron mobility in the 2DEG is
typically $ 4 - 6
\times 10^6$ cm$^2$V$^{-1}$s$^{-1}$ at 4.2K and increases
further at lower temperatures. At 100mK the 2D ballistic mean free paths exceed
$160 \mu m$ \cite{2D_PRB}  which is greater than our sample dimensions. These devices 
are comparable with the highest mobility electron systems yet produced.
Ballistic conductance plateaus have been demonstrated in quantum wires up
to $5 \mu m$ in length, with the data exhibiting more than 15 plateaus \cite{BK_QWAPL}. 

To investigate the sensitivity of many-body effects to the length of the
1D region, we
have measured the conductance of
quantum wires of nominal length $l$=$0$, $2 \mu m $ and  $5 \mu m $.
The devices were patterned from  ultra-high-mobility heterostructures,
comprising a 75 nm  layer
of Al$_{0.3}$Ga$_{0.7}$As on top of GaAs to produce the 2DEG interface. A 25 nm GaAs
spacer separated the epitaxial conducting
top gate from the AlGaAs. NiAuGe ohmic contacts were made to the 2DEG using a
self-aligned technique.
Electron beam
lithography and shallow wet etching were used to selectively remove the top
gate to form  the
quantum wires. The top gate was sectioned into three
separately controllable gates.
The center (top gate) was biased positively relative to the contacts to
induce a 2DEG at the
GaAs/AlGaAs interface. This
positive bias $V_T$ determined the carrier density in the
2DEG reservoirs which was tunable from $0.6 - 6 \times 10^{11}$cm$^{-2}$ corresponding to $V_T$=0.05V to 0.8V.
 A negative voltage $V_s$ was  then applied to the side
gates to produce
electrostatic 1D confinement in addition to the geometric confinement
already present (see Figure 1). 

Low frequency four-terminal conductance
measurements were made with an excitation voltage
below  $10 \mu V$
using two lock-in amplifiers to monitor both current and voltage. We
stress that the results presented here are
raw data as no equivalent series resistance has been subtracted and no attempt has
been made to adjust the
plateau heights to fit with quantised units of $ 2e^2/h$.  

\small
\centerline{$\phantom{X}$} \par\noindent%
\leftline{\psfig{figure=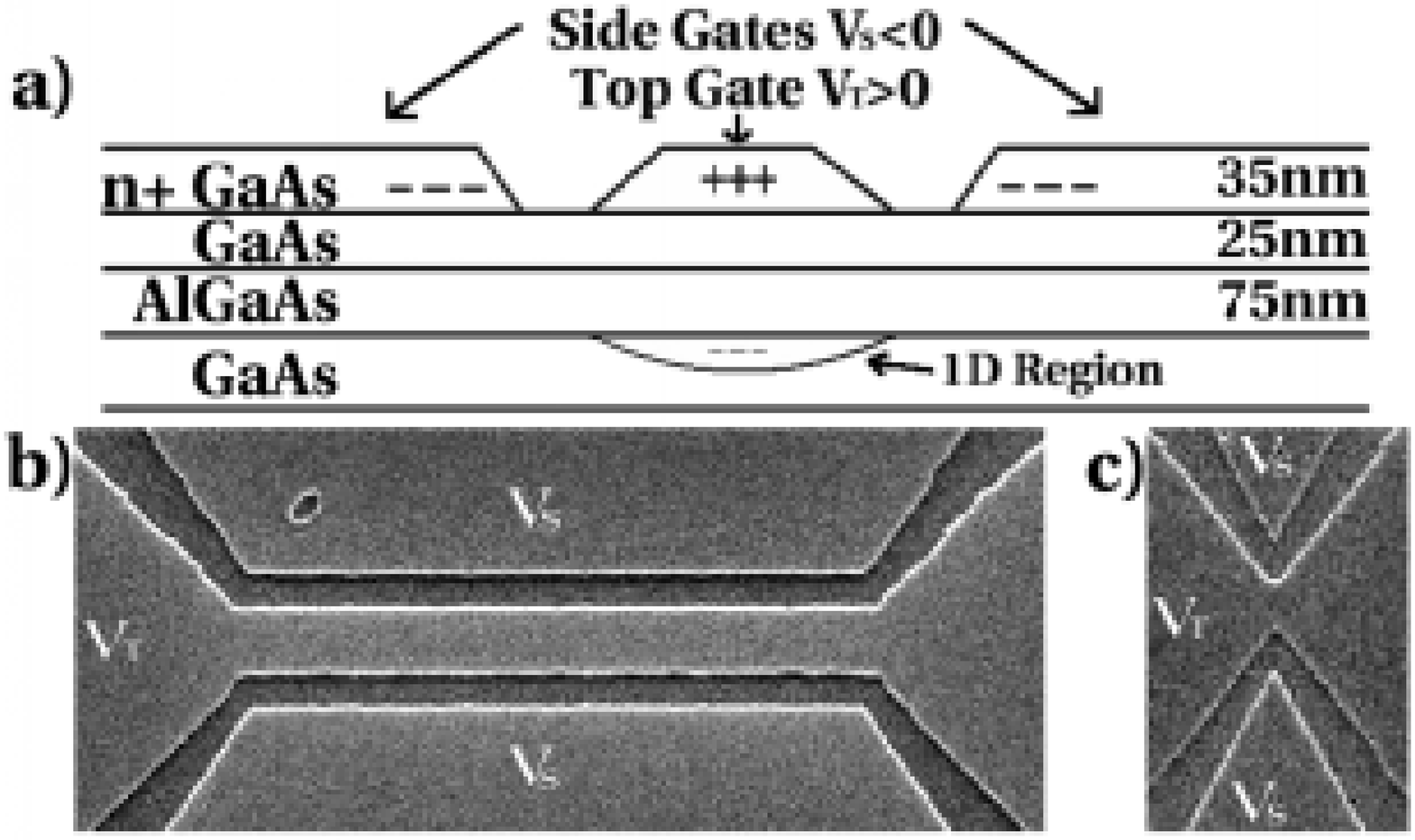,width=8.6cm,height=5.6cm}} 
\noindent%
{\bf Figure 1:} a) Cross-sectional schematic of a quantum wire, showing the positively biased top
gate and the side gates biased negatively. b) SEM micrograph of a quantum wire with length $l=5\mu m$. c) SEM
micrograph of a quantum wire with length $l=0\mu m$.
\centerline{$\phantom{X}$}
\normalsize

The conductance $G$ of a zero-length quantum wire  is shown in Figure 2  as a
function of the side gate voltage
$V_s$ at a temperature {\em T} = 50mK.  Data were taken at a series of top gate voltages  corresponding
to different 1D densities. The 1D electron density  $n_{1D}$
may be controlled using {\em both} the top and side gates to vary the
shape of the  potential well perpendicular to the channel.
 When both the top and side gates are
strongly ({\em weakly}) biased positive and negative respectively the confining potential is steep
({\em shallow}),
leading to a large ({\em small}) 1D sub-band spacing and a corresponding high ({\em low}) 1D electron
density.
In
this way it is possible to maintain a constant 1D occupancy, and hence conductance, while varying $n_{1D}$.

For $G <  2e^2/h$
an additional feature is  observed close to  $0.7 \times 2e^2/h$, as seen by others
\cite{Thomas_spin1st,Thomas_int,BK_QWAPL,Kristensen,Tsch1}. A similar feature is
also observed in a
second identical quantum wire with length $l=0$ (not shown). 
As with other workers we find that this feature is robust to cryogenic cycling, indicating that it is 
unlikely to be related to an impurity state.
Although the data in Figure
2 implies a small enhancement of the  $0.7$ structure with increasing $n_{1D}$, the
trend is not fully monotonic and is less so for the second $l=0$
wire
measured. The inset of
Figure 2 shows the
temperature dependence of the conductance for the wire with length $l=0$. These
temperature measurements were made with an average electron density $(n_{2D}
\approx 3 \times 10^{11}$cm$^{-2}$, $V_T =
0.4$V). Similar behavior with temperature is seen at high
and low
electron densities in both of the  zero length quantum wires studied. 
This temperature dependence deviates 
from the expected  single particle result with little thermal smearing below $0.7 \times 2e^2/h$.    
Such puzzling behavior is consistent with measurements made by
others \cite{Thomas_spin1st,Kristensen}. \\

\small
\centerline{$\phantom{X}$} \par\noindent%
\centerline{\psfig{figure=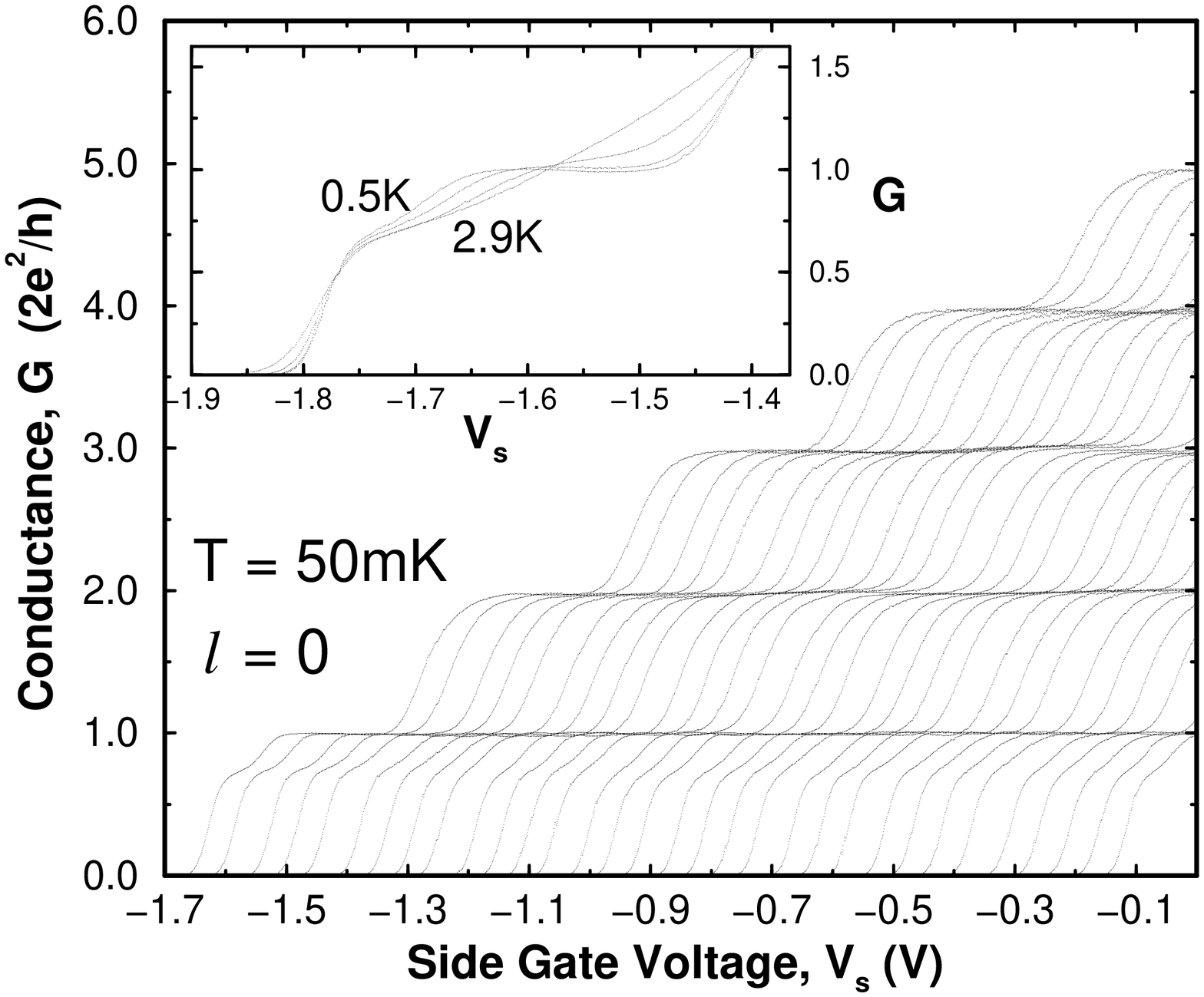,width=8.2cm,height=6.8cm}}
\noindent%
{\bf Figure 2:} Conductance measurements of a $l=0$ 
quantum wire as a function of
side gate voltage for
top gate voltages, $V_T$ = 172mV - 300mV (right to left) in steps of 4mV. 
Inset: Temperature
dependence of the conductance at $V_T$=0.4V.
The curves
are for temperatures 0.5K,
1.0K, 1.5K \& 2.9K. 
\centerline{$\phantom{X}$} \par\noindent%
\normalsize

The results in Figure 2 demonstrate that these epitaxially gated nanostructures produce 
ultra-low-disorder quantum wires for $l=0$ which exhibit the $0.7 \times 2e^2/h$ conductance feature
comparable with 
the strongest so far observed. When we extend to longer quantum wires, new and unexpected
results are seen. 

Figure 3 shows the conductance {\em G} of a quantum wire with $l=2\mu m$ 
as a function of side gate voltage $V_S$. The density $n_{1D}$ increases 
from right to left as the confining potential is steepened. Data were obtained at temperatures $T$=1K and $T$=50mK. Clear
conductance
quantisation is seen near integer multiples of $2e^2/h$ with up to 15 platueas evident, 
indicating ballistic transport along the full length of the $2\mu m$ wire, as previously reported 
\cite{BK_QWAPL}. 
The data collected at $T$=1K show a clear plateau-like feature which becomes more pronounced and 
evolves downwards in $G$ towards
$0.5 \times 2e^2/h$ as $V_T$ is increased. A much weaker inflection is also present near $0.7 \times 2e^2/h$ 
across the full density range.
Further evidence of many-body phenomena is seen evolving  
in the range  $1.5 - 1.7 \times 2e^2/h$ where the structure is strong enough to resemble the conductance feature
seen near $ 0.7 \times 2e^2/h$ in quantum  wires with $l=0$.  \\

\small
\centerline{$\phantom{X}$} \par\noindent%
\centerline{\psfig{figure=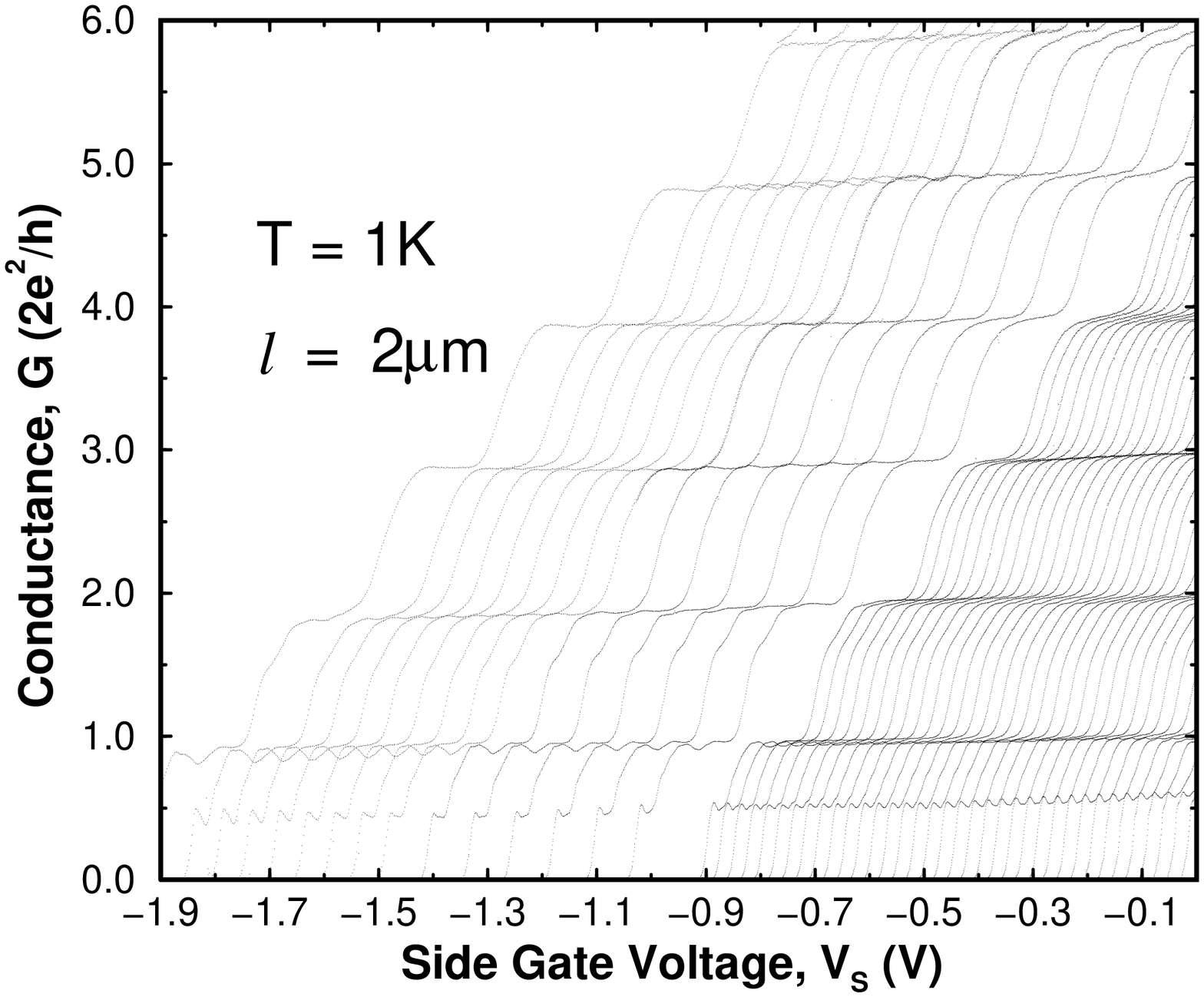,width=8.2cm,height=6.8cm}} \\

\centerline{\psfig{figure=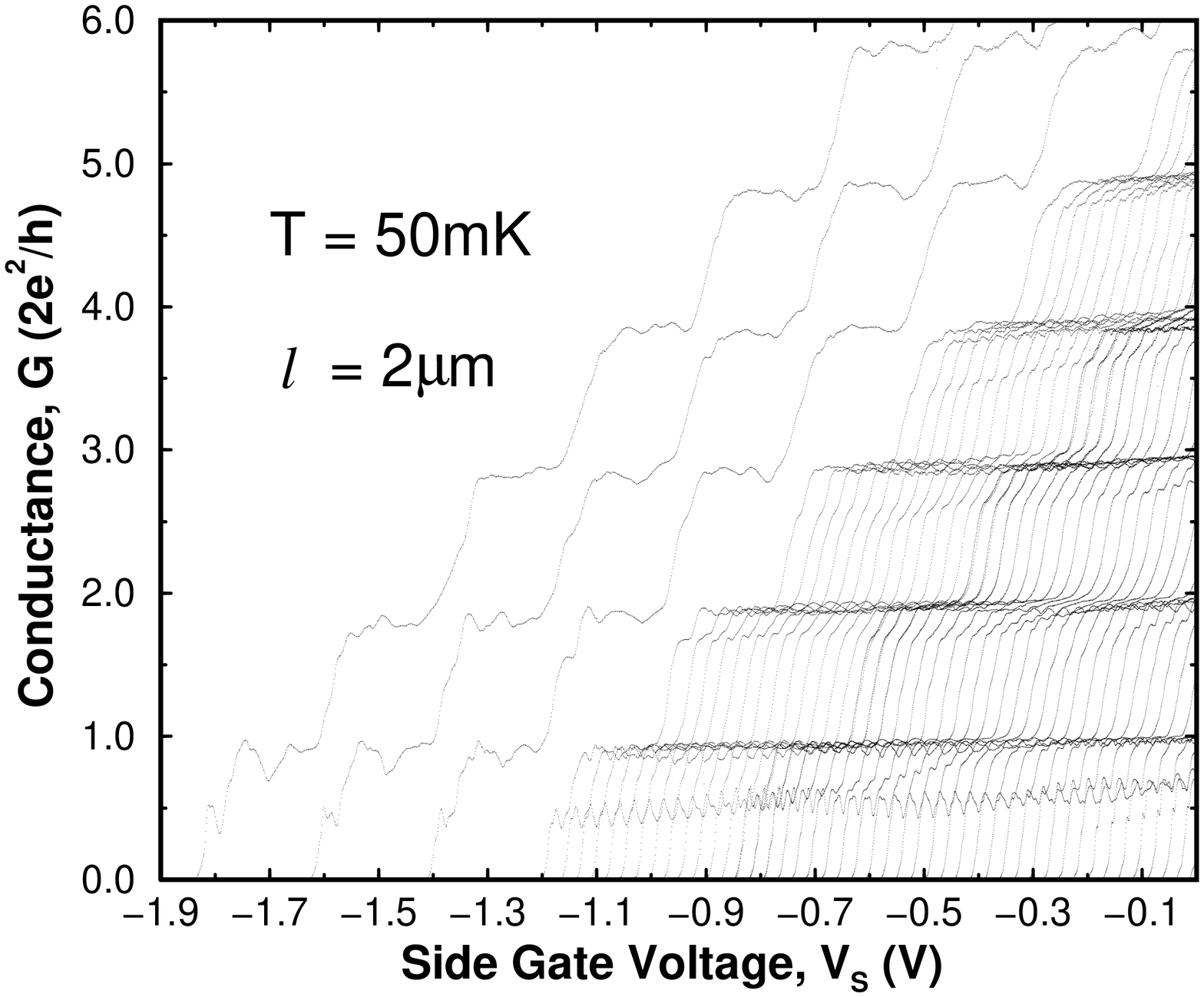,width=8.2cm,height=6.8cm}}
\noindent%
{\bf Figure 3:} Conductance of a $2 \mu m$ quantum wire. Conductance as a function of
side gate voltage for top gate voltages, $V_T$ = 300mV - 620mV (right to left). The data at the top was
obtained close to 1K and the bottom section was obtained at approximately 50mK.
\centerline{$\phantom{X}$} \par\noindent%
\normalsize

Conductance measurements of quantum wires with $l=5\mu m$
exhibit similar
plateau-like
features near $0.5 \times 2e^2/h$ to wires with $l$=$2\mu m$, as noted in our previous work \cite{BK_QWAPL}. 
However, for  $l$=$5\mu m$
the weak disorder which is present leads to a distortion of the conductance plateaus, making interpretation more 
difficult and here we focus on wires with $l=2 \mu m$ where the
single particle plateaus are as clear as those seen in $l=0$ devices. 

As the  $2 \mu m$ wire is cooled to $T$ = 50mK the feature near $0.5 \times 2e^2/h$ remains, however, rich evolving 
structure is also revealed. 
Conductance inflections occur below each of the integer plateaus (within $e^2/h$) which predominantly evolve downwards in $G$
with increasing $n_{1D}$. 
One explanation within a single-particle picture is weak disorder, leading to interference 
of electron waves along the quantum wire. 
However, against this, remnants of the strongest features survive 
at $T$=1K, in particular the feature below $2 \times 2e^2/h$ is reminiscent of the  $0.7 \times 2e^2/h$ 
feature seen in low-disorder $l$=0 wires, implying a possible many-body origin.  \\

\small
\centerline{$\phantom{X}$} \par\noindent%
\centerline{\psfig{figure=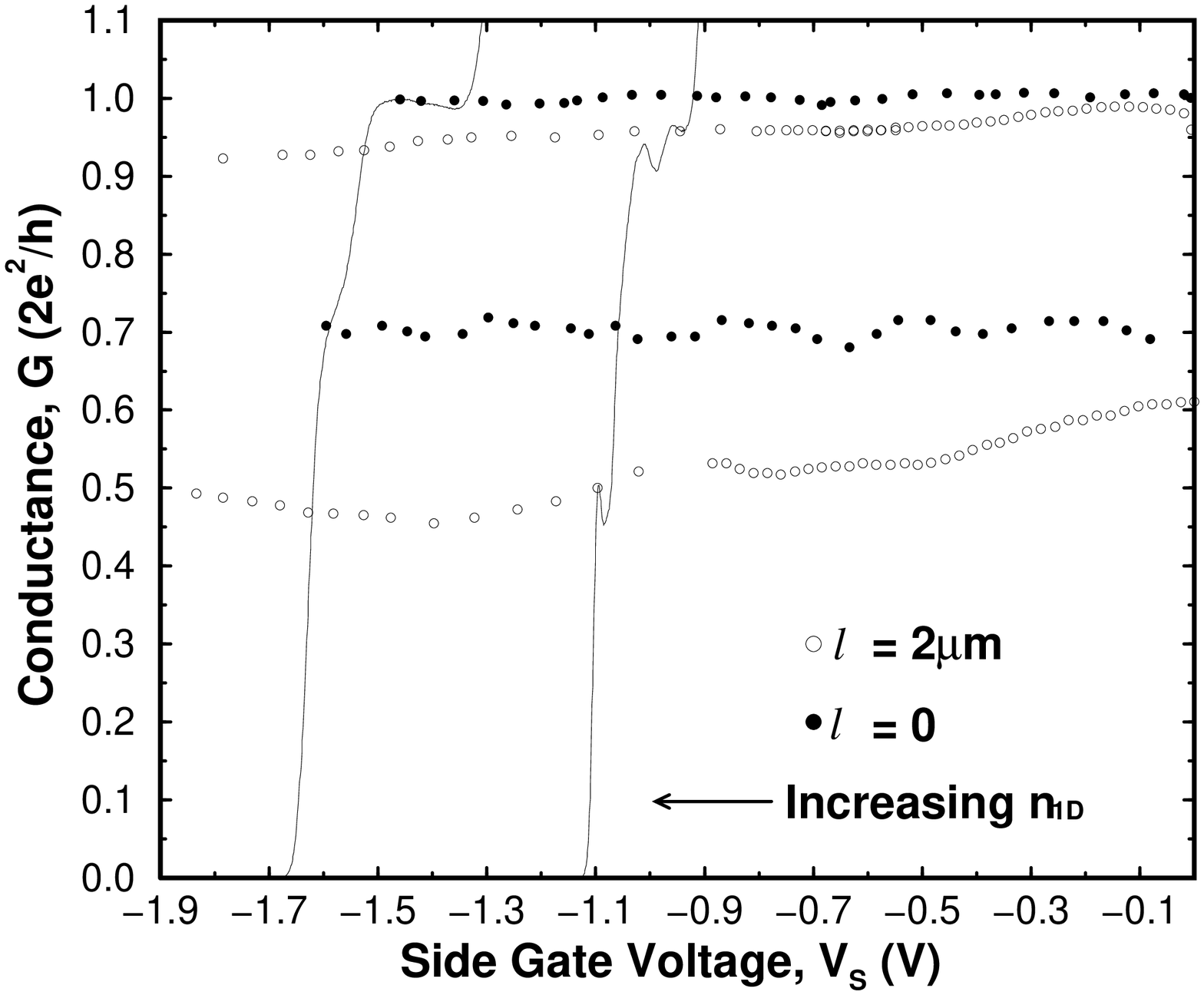,width=8.2cm,height=6.8cm}}
\noindent%
{\bf Figure 4:} Evolution of the conductance features seen in $l=0$ and $2 \mu m$
quantum wires as a function of $n_{1D}$. The evolution of the corresponding plateau
at  $2e^2/h$ is also shown. Single conductance traces are included for
both devices to facilitate interpretation. Open circles are  $2 \mu m$ data
and closed circles are $l=0$ data.
\centerline{$\phantom{X}$} \par\noindent%
\normalsize

Figure 4 details the evolution of the conductance features seen in quantum wires of length
$l=0$ and $2\mu m$ with varying
$n_{1D}.$ We define the position of the  feature seen in $l=0$ devices near
$0.7 \times 2e^2/h$ as the conductance $G$ at which $dG/dV_s$ is a local minimum. In a similar manner, 
for the $l=2
\mu m$ quantum
wire we define the position of the plateau-like feature  as the first 
local minimum in the  $dG/dV_s$
curve for $T$=1K.

Note that the plateau
at $2e^2/h$ remains almost constant for the $l=0$ wire but for $l=2\mu m$ the plateau falls in $G$ (by up to 8\%) as $n_{1D}$ is
increased.  Suppression of plateaus below the ideal quantised values has been observed in 
previous studies on quantum wires  \cite{Tarucha_quant,Yacoby_QWR,BK_QWAPL}, 
and considered theoretically in a number of many-body treatments \cite{KandF_LLPRL,Maslov_endNov}. In our case the suppression cannot be
explained by a simple increase of the
effective
series resistance associated with the 2D contact regions, since
 the 2D sheet resistance decreases
with increasing $V_T$.
Abrupt coupling of the 2D  reservoirs to the low
density 1D region could result in a reduction of the transmission coefficient as the 2D electron density is increased. 
We note that the density mismatch is larger for the
longer wire, since the top-gate voltage threshold for conduction is almost twice as large for $l=2\mu m$ as for $l=0$.

Turning now to the non-integer plateaus we see that the feature near $0.7 \times 2e^2/h$  
in $l=0$ devices becomes slightly  more {\em pronounced} with increasing $n_{1D}$ (Figure 2)
but the variation in conductance is small. This is in contrast to the plateau-like feature seen in 
the $l=2 \mu m$ wire data, which evolves towards $0.5 \times 2e^2/h$ with increasing $n_{1D}$. It is significant 
that if the single particle plateau  for $l=2\mu m$ is normalised to equal $2e^2/h$
then this feature still evolves downwards in $G$ but  never falls below  $0.5 \times 2e^2/h$, the position expected for a spin-split
1D plateau.

Conductance data suggestive of many-body effects in 1D have now  been observed in a variety of high mobility structures including 
split-gated HEMTs 
\cite{Thomas_spin1st}, gate metallised structures \cite{Kristensen} and our undoped enhancement mode FETs considered here. Some evidence
for this effect has also
been seen in low mobility quantum wires based on ion-beam defined GaAs transistors \cite{Tsch1} and other material systems
such as GaInAs/InP \cite{ramvall} and 
 n-PbTe  \cite{grabecki}.  The diverse number of experimental systems that have  been examined 
would seem to establish the feature  as an intrinsic property 
of a 1D correlated system. In particular the
temperature dependence, 
described as {\em activated} by \cite{Kristensen} and  
detailed in our $l=0$ wires, remains consistent between devices of different design \cite{Thomas_spin1st,Kristensen}. 
Some important exceptions do
exist, however, as in measurements of narrow wires by Yacoby {\em et al.} \cite{Yacoby_QWR}  
and Tarucha {\em et al.} \cite{Tarucha_quant} there appears to be no
{\em strong}
feature present even though clear
quantisation is seen. The absence of the 
feature in  reference \cite{Yacoby_QWR} may be  associated with a large 1D sub-band spacing made possible 
in that case due to a novel epitaxial confinement technique. 
 
The most commonly invoked explanation for additional conductance structure near $0.7 \times 2e^2/h$ has been some form of spontaneous
spin polarisation mediated through the exchange interaction, as detailed in references \cite{Gold_gpar,wang}. The possibility of a ferromagnetic
instability below a critical electron concentration has also been considered \cite{byczuk}. It has so far remained a mystery as to why 
measurements show structure near $0.7 \times 2e^2/h$, rather than  $0.5 \times 2e^2/h$, the value expected for a fully 
spin-polarised 1D level. The fact
that we see structure near  $0.7 \times 2e^2/h$ in $l=0$ wires and structure evolving towards  $0.5 \times 2e^2/h$ in longer 
wires (with $l=2\mu m$ and $l=5\mu m$
\cite{BK_QWAPL}) leads to a possible scenario in which spin-splitting is only fully resolved in wires above some critical length scale. 
The additional structure we observe
near  $1.7 \times 2e^2/h$, and in higher sub-bands below 1K, also 
suggest that many-body effects become enhanced in longer 1D regions. We note that conductance anomalies in higher
sub-bands have been predicted in reference \cite{wang}.

An alternative explanation for the $0.7 \times  2e^2/h$ feature has been argued in two recent theories \cite{flambaum,coloumb} which
consider a scenario in which two- (or more) body processes dominate. In the proposed case where 
electron pairs dominate transport and $l=0$, the three
triplet states are lower in energy than the one singlet state, leading to a plateau 
at (slightly less than)  $0.75 \times 2e^2/h$, since the triplet states are
transmitted (with transmission coefficient not precisely 1) while
the singlet is reflected by the constriction. Within this model the observation in our $l=2\mu m$ 
wire of a dominant feature  near  $0.5 \times 2e^2/h$ together with remanent weaker structure
in the vicinity of  $0.75 \times 2e^2/h$, which is still noticeable at $T$=1K (see Figure 3),
would be interpreted as the dominance of spin polarisation in the finite length wire with increasing $n_{1D}$ 
with some remanent signature associated with the singlet/triplet mechanism. However this 
observation could also be compatible with 1D Wigner crystallisation (dominant
repulsive interaction) 
providing the Landauer-B\"{u}ttiker framework
can be extended to this regime \cite{flambaum}. 

In conclusion, we have studied ultra-low-disorder quantum wires utilising a novel GaAs/AlGaAs  
layer structure which avoids the 
random impurity potential associated with modulation doping, making these devices ideal for the 
study of electron correlations in
1D. In common with other workers we find structure near  $0.7 \times 2e^2/h$ in wires with $l=0$, whereas in 
longer wires the dominant structure evolves
towards  $0.5 \times 2e^2/h$ at high 1D carrier concentrations. It is not possible to be conclusive as to whether 
these effects are related to a
many-electron spin polarisation, 
or to a more complex explanation (for example 1D Wigner crystalisation). However it is clear that both the length over which
interactions occur and the 1D density play an important role in determining the effect of these
mechanisms upon electrical transport.

We are grateful to B. Altshuler, V. Flambaum and  A. R. Hamilton for valuable discussions and 
R. P. Starrett for technical support. This work was funded by the Australian Research
Council grant A69700583.

\end{document}